\documentclass{article}
\RequirePackage[a4paper]{geometry}
\usepackage{amsmath,graphicx}

\usepackage{tikz}

\usepackage{placeins}  %
\usepackage[utf8]{inputenc}

\usepackage{xcolor}
\definecolor{blue}{HTML}{1F77B4}
\definecolor{red}{HTML}{D62728}
\definecolor{green}{HTML}{2CA02C}
\definecolor{orange}{HTML}{FF7F0E}

\usepackage[
    colorlinks=true,
    linkcolor=black,
    citecolor=black,
    filecolor=black,
    urlcolor=black,
    plainpages=false,
    pdfpagelabels,
    breaklinks=true,
    pdftitle={Gradient-based methods for spiking physical systems},
    pdfsubject={},
    pdfauthor={J. Goeltz, S. Billaudelle, L. Kriener, L. Blessing, C. Pehle, E. Mueller, J. Schemmel, M. A. Petrovici},
]{hyperref}

\usepackage[
    citestyle=numeric-comp,
    doi=false,
    giveninits=true,
    isbn=false,
    maxbibnames=3,
    minbibnames=2,
    sorting=none,
]{biblatex}
\AtEveryBibitem{\clearfield{eprint}}
\AtEveryBibitem{\clearfield{pages}}
\AtEveryBibitem{\clearfield{url}}
\AtEveryBibitem{\clearlist{address}}
\AtEveryBibitem{\clearlist{location}}
\AtEveryBibitem{\clearlist{series}}

\usepackage{tabularx}
\newcolumntype{Y}{>{\centering\arraybackslash}X}

\usepackage{tabularray}
\usepackage{todonotes}

\usepackage{changes}

\usepackage{tikz}

\usepackage{glossaries-extra}
\setabbreviationstyle[acronym]{long-short}
\glssetcategoryattribute{acronym}{nohyperfirst}{true}

\newacronym{snn}{SNN}{spiking neural network}
\newacronym{bptt}{BPTT}{backpropagation through time}
\newacronym{lif}{LIF}{leaky integrate-and-fire}

\begin{document}
\renewcommand{\abstractname}{\vspace{-\baselineskip}}

\begin{center}
{\Large\bfseries
    Gradient-based methods for spiking physical systems
\par}
\vspace{3ex}

\noindent
{\bfseries
J. Göltz$^{\star1,2}$, S. Billaudelle$^{\star1}$, L. Kriener$^{\star2}$, L. Blessing$^1$, C. Pehle$^1$, 
\\
E. Müller$^1$, J. Schemmel$^1$, M. A. Petrovici$^2$
\par
}
{\footnotesize\itshape
$^\star$equal contributions
$^1$Kirchhoff-Insitute for Physics, Heidelberg University
$^2$Department of Physiology, University of Bern
\par}
\vspace{0ex}
\end{center}

\begin{abstract}
    \noindent \textbf{Summary}
    Recent efforts have fostered significant progress towards deep learning in spiking networks, both theoretical and in silico.
    Here, we discuss several different approaches, including a tentative comparison of the results on BrainScaleS-2, and hint towards future such comparative studies.
\end{abstract}

\FloatBarrier

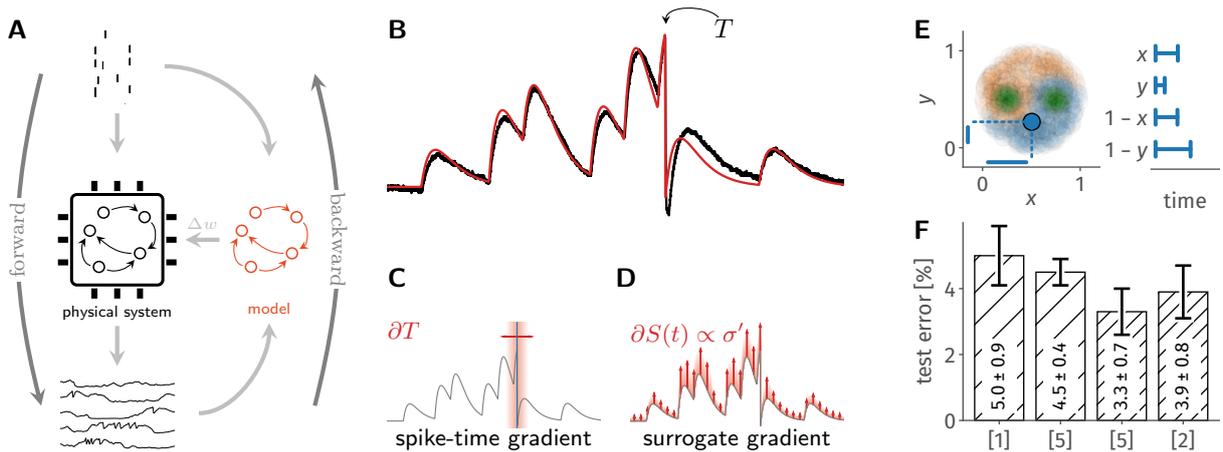
\begin{figure}[!b]
	\begin{centering}
        \vspace{-0.5em}
        \begin{tikzpicture}
            \tikzset{panel/.style={inner sep=0pt}};
            \tikzset{panellabel/.style={anchor=north west, inner sep=0pt, font={\bfseries\sffamily}}};
            \node[panel, anchor=north west] (a) at (0, 0) {
                    \includegraphics{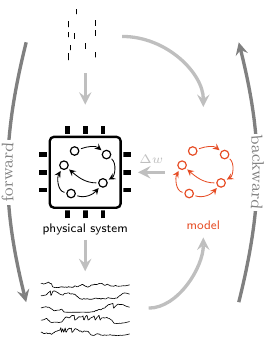}
            };
            \node[panel, anchor=north west] (b) at (5, 0) {
                \begin{tikzpicture}
                    \node (trace) at (0, 0) {\includegraphics[width=6cm]{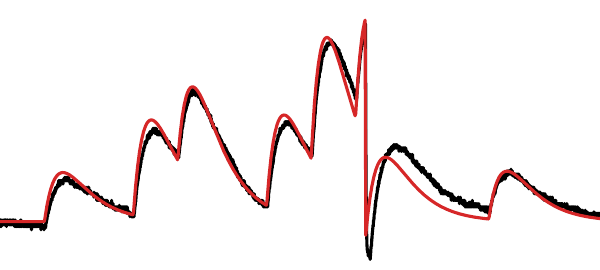}};

                    \node[anchor=north west, yshift=-1.3cm] (fnd) at (trace.south west) {\includegraphics[width=2.8cm]{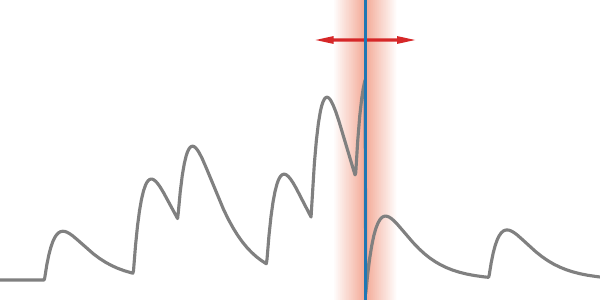}};
                    \node[anchor=north east, yshift=-1.3cm] (sgd) at (trace.south east) {\includegraphics[width=2.8cm]{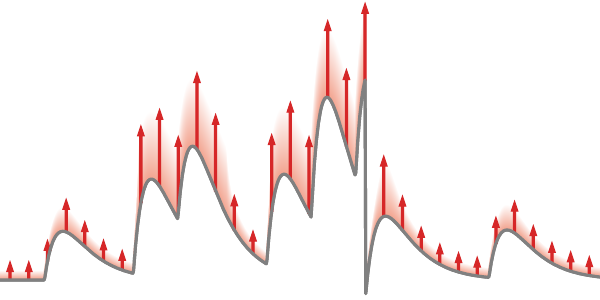}};
                    \node[below] at (fnd.south) {\small\sffamily spike-time gradient};
                    \node[below] at (sgd.south) {\small\sffamily surrogate gradient};

                    \node[red, inner sep=0pt, anchor=north west] at (sgd.north west) {\small $\partial S(t) \propto \sigma^\prime $};
                    \node[red, inner sep=0pt, anchor=north west] at (fnd.north west) {\small $\partial T$};
                \end{tikzpicture}
            };
            \node[panellabel] (spiketimeLabel) at ([yshift=0.0cm,xshift=4.3cm]b.north west) {$T$};  %
            \coordinate (spiketime) at ([yshift=-0.1cm, xshift=3.67cm]b.north west);
			\draw[-stealth, shorten >=1pt, shorten <=1pt] (spiketimeLabel.110) to[out=135, in=90] (spiketime);
            \node[panel, anchor=north east] (e) at (\textwidth, 0) {
                \includegraphics[width=1.6in]{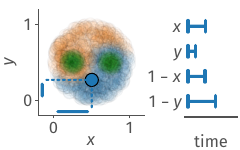}
            };
            \node[panel, anchor=south east] (f) at (b.south -| e.east) {
                \includegraphics[width=1.6in]{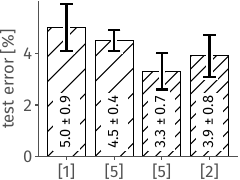}
            };
            \node[panellabel] at (a.north west) {A};
            \node[panellabel] at (b.north west) {B};
            \node[panellabel] at (e.north west) {E};
            \node[panellabel] at (f.north west) {F};
            \node[panellabel] at ([yshift=-3.3cm]b.north west) {C};
            \node[panellabel] at ([yshift=-3.3cm, xshift=3.0cm]b.north west) {D};
        \end{tikzpicture}
		\par\end{centering}
         \vspace{-1em}
          \caption{
            \small
            \textbf{A}
                In-the-loop approaches evaluate the physical system in the forward pass based on input spikes (raster plot, top).
                Combined with recorded observables (e.g., voltages and spikes, bottom), a model of the internal dynamics is then used to estimate gradients for weight update calculations.
            \textbf{B}
                Example data recorded from BrainScaleS-2 (black), as well as a simulated model trace replicating the core dynamics of the silicon neuron (red).
            \textbf{C}
                Spike-time gradients \cite{goeltz2021fast,pehle2023event} calculate the spike times' derivatives $\partial T$ w.r.t. changes in presynaptic spike times or afferent weights.
                This quantity is only defined at the spike times and thus extremely sparse in time.
            \textbf{D}
                Surrogate gradients \cite{neftci2019surrogate} estimate how a parameter change would affect the `likeliness' of a spike at each point in time by replacing the `hard' threshold $\theta$ by the smooth surrogate $\sigma$.
            \textbf{E}
                The Yin-Yang data set \cite{kriener2021yin} and a corresponding temporal encoding of a randomly chosen sample.
            \textbf{F}
                Comparison of the three methods~\cites[]{goeltz2021fast,pehle2023event}[new results for][]{cramer2022surrogate}.
                Sparse and dense hatching indicate time-to-first-spike and voltage-based output decoding, respectively.
            \vspace{-1em}
          }\label{fig:sim_DS}
\end{figure}

\vspace{-1.1em}
\paragraph{Introduction}
Physical computation directly exploits the intrinsic dynamics of a given substrate to efficiently process and propagate information.
In contrast to numerical computers, physical computers implicitly obey the dynamics required by certain models of information processing (e.g., neuronal integration) rather than calculating them explicitly by arithmetically manipulating binary representations thereof.
Neuromorphic computers represent a prominent class of physical systems, drawing inspiration from the nervous system by mimicking the dynamics of neurons and synapses.
They typically involve a massively parallel and time-continuous implementation of neuro-synaptic dynamics as well as an asynchronous event-based propagation of signals to efficiently emulate \glspl{snn}.

Physical systems are typically ``programmed'' by tuning their internal dynamics, such as time constants or coupling strengths.
In our case, we employ gradient-based optimization schemes to adapt the emulated \glspl{snn} to a given task.
Here, we discuss multiple approaches which all have been demonstrated on the mixed-signal neuromorphic system BrainScaleS-2~\cite{pehle2022brainscales2}.
With this demonstration they have proven to address both the problems arising from the event-based characteristics of \glspl{snn} in general and the analog nature of the substrate in particular.

\vspace{-1.1em}
\paragraph{In-the-loop training of physical systems}
With self-adapting, local neuromorphic learning still in its infancy,
we sometimes take inspiration from machine learning when training our neuromorphic devices, in particular from gradient-based optimization.
To adopt these for physical computation, we need \emph{differentiable estimates} of the system-internal dynamics.
In our case of a time-continuous spiking neuromorphic system, this \emph{model} has to capture the propagation and weighting of spikes as well as the neuronal dynamics.
This could be realized as a collection of closed-form expressions for spike times or through a computation graph describing the propagation of stimuli through the network.
An exact match between model and system dynamics is, however, often unattainable, and in a trade-off between model fidelity and complexity, higher-order effects are typically neglected.
Instead, a continuous synchronization between model and system can mitigate the propagation of incorrect estimates.
For this purpose, gradients are estimated based on \emph{measurements of observables}, including spike times or even membrane potential traces.
Generally speaking, precise knowledge of the internal state can often be traded against model fidelity and vice versa.

\vspace{-1.1em}
\paragraph{Gradient calculation}
While the time-continuous and highly non-linear nature of \glspl{snn} impedes a straight-forward gradient descent, there exist multiple approaches for the estimation of gradients.
These can be coarsely divided into methods either involving exact derivatives or resorting to inherent approximations of the spike triggering threshold.

\textit{Exact, sparse gradients} can be obtained for a \gls{lif} neuron's output spike times with respect to both input weights and presynaptic spike times.
Analytical expressions for the gradients can -- under certain assumptions -- be derived based on differentiable  expression for the spike time $T$ as a function of only the input spikes and weights $T(\{t_i\}, \{w_i\})$~\cite{goeltz2021fast}.
The derivatives of this function allow assignment of credit through multiple layers, and consequently gradient descent in deep networks.
At the expense of an analytical solution but relaxing some assumptions, gradients can also be computed by relying on a backward evolution of adjoint dynamics~\cite{wunderlich2021event}.
Both approaches give exact relations on how to change weights to shift the \emph{existing} spikes to reduce the loss function.

\textit{Surrogate gradients} \cite{neftci2019surrogate} offer an alternative approach by considering the output spike train $S(t) = \sum_i \delta(t-T_i)$ of a neuron, where the individual spike times are given by $T_i$.
In order to estimate useful gradients not only at the individual output spike times, $\partial S(t)$ can be \emph{approximated} with the help of a surrogate derivative $\sigma^\prime(v(t))$ based on the membrane potential $v(t)$.
In contrast to exact, spike-time-based approaches, this method also assigns gradients at times where no spikes occur.
While leading to a potential memory and computation overhead, this approach  allows explicit awareness of the creation or deletion of spikes and can hence innately train networks from a quiescent state by specifically recruit neuronal activity where required.

\vspace{-1.1em}
\paragraph{Results}
The three approaches outlined above have all been demonstrated on the BrainScaleS-2 system.
Figure \ref{fig:sim_DS}F shows respective results obtained for the Yin-Yang dataset \cite{kriener2021yin}.
All three methods are able to successfully solve the task and yield comparable classification accuracies.
Classifiers based on a time-to-first-spike output seem to incur a slight performance penalty in comparison to voltage-based outputs.
Further, approaches based on exact spike-time gradients appear to yield accuracies slightly below the ones reached by surrogate gradient methods.
However, inhomogeneous choices of system- and hyperparameters spoil a direct comparison across publications.
In addition to the presented results, BrainScaleS-2 has been trained on a variety of other datasets, with both feedforward and recurrent network topologies~\cite{goeltz2021fast,cramer2022surrogate}.
The respective training methods were shown to be robust against parameter noise that was artificially induced to mimic fixed-pattern deviations not uncommon in physical systems.

\vspace{-1.1em}
\paragraph{Discussion}
The capability to train highly complex, nonlinear dynamical systems, not just in idealized simulations, but also in practice, represents a fundamental prerequisite for the real-world deployment of neuromorphic devices.
Here, we have reviewed three recently demonstrated methods and their results on BrainScaleS-2~\cite{cramer2022surrogate,goeltz2021fast,pehle2023event}.
This demonstration is a testament to the maturity of this system in particular as well as to the progress of physical computation in general.
It also paves the way for important future studies including comparisons on a variety of performance metrics such as
convergence speed during training, data efficiency, robustness to noise and parameter changes.

\vspace{-0.8em}
{\small
\paragraph{\small Acknowledgements}
This work received funding from the EU research and innovation funding H2020 945539 (HBP SGA3), DFG project EXC 2181/1390900948 (STRUCTURES), and the Manfred Stärk Foundation.
}

\vspace{-0.9em}
\paragraph{\small References}{
}
\AtNextBibliography{\small}
\begingroup
\setlength\bibitemsep{0pt}
\printbibliography[heading=none]{}
\endgroup

\end{document}